\documentclass{elsart}

\usepackage{amssymb}
\usepackage{graphicx}

\begin{document}

\begin{frontmatter}

\title{The Spectrometer System for Measuring ZEUS Luminosity at HERA}

\author{M.~Helbich, Y.~Ning, S.~Paganis, Z.~Ren, W.B.~Schmidke, F.~Sciulli}
\address{\it Nevis Laboratories, Columbia University, Irvington on Hudson, 
New York}

\author{U.~Schneekloth}
\address{\it Deutsches Elektronen-Synchrotron DESY, Hamburg, Germany}

\author{C.~B\"uttner, A.~Caldwell, J.~Sutiak}
\address{\it Max-Planck-Institut f\"ur Physik, M\"unchen, Germany}

\begin{abstract}
The upgrade of the HERA accelerator has
provided much increased collider luminosity.  In turn, the improvements have
necessitated a new design for the ZEUS luminosity measurements.
The intense synchrotron radiation field, as well as the high probability of a
bremsstrahlung photon in each bunch crossing, posed new experimental
constraints. In this report, we describe how these challenges were met
with the ZEUS luminosity spectrometer system.  The design, testing and
commissioning of the device are described, and the results from the
initial operational experience are reported.
\end{abstract}

\end{frontmatter}

\journal{NIM in Physics Research, Section A}

\section{Introduction}
\label{sec:introdcution}
ZEUS is one of two detectors accumulating data from electron-proton beam 
collisions at the HERA accelerator operated by the DESY laboratory in 
Hamburg, Germany.  Analyses of such data have provided quantitative values 
of cross sections and derivative information related to the proton's quark 
and gluon content, Quantum Chromodynamics, and other issues relevant to the 
Standard Model and beyond.  In September 2000, HERA completed an eight year 
running period (HERA-I), which provided measurements of deep inelastic 
scattering (DIS), photoproduction, and other processes in a newly accessible 
kinematic 
regime. HERA underwent a luminosity upgrade in 2001 \cite{hera-98-05}.  
The goal was to increase the HERA-I peak luminosity by a factor of 5.
In addition to the accelerator modifications necessary to achieve this, 
spin rotators were installed to provide electron beam longitudinal 
polarization of about $40\%$ on average.  
A new physics program (HERA-II) began in 2002 
to make accurate measurements of small cross sections with polarized beams.

Precise knowledge of the luminosity is required for precise determination of 
a cross section associated with any process; such measurements depend on 
luminosity integrated over time, $\int Ldt$, to normalize the numbers 
of events observed during the same time period.  
Luminosity had been measured \cite{np:b316:269,acpp:b32:2025,misc:ichep96:h1} 
in HERA-I using the rate of high energy photons from the bremsstrahlung 
process, $ep \rightarrow ep\gamma$,
produced by the colliding beams.  This process is well understood, has a high
rate and an 
accurately calculable cross section.  Produced photons follow the direction 
of the colliding beam electrons and are observed about 100~m downstream, 
after the electron and proton beams have been magnetically separated.  
The ZEUS HERA-I technique, with a calorimeter to directly measure all 
bremsstrahlung photons faces new difficulties at HERA-II, 
including the following:
\begin{itemize}
\item
A significant increase in direct synchrotron radiation (SR) flux 
from the electron beam occurs due to the higher beam currents and to a new 
beam focusing scheme to optimize high luminosity near the interaction region.
\item
Much larger numbers of overlaid bremsstrahlung events (pile-up) occur.  
Colliding bunches producing more than one photon are not separated in the 
calorimeter technique.  HERA-II luminosity implies a significant 
probability (approaching unity) for several final state photons with
$E_{\gamma} > 0.5$~GeV 
in each bunch-crossing.  
\item
There are additional requirements for accurate cross sections 
using polarized beam electrons (or positrons).
\end{itemize}

The HERA beams each consist of 220 bunches separated by 96~ns. A few bunches 
are unfilled; the empty bunches are used for monitoring purposes and they
are also needed for the proton beam abort system, and the 
unpaired bunches are used to measure small backgrounds resulting from the 
electron or proton beam interacting with residual gas near the 
intersection region.  
After acceleration, the polarization of the electron beam 
\cite{upub:polarization2000:polarization2000} 
rises from $P=0$ to $P=50\%$ over a period of about $\tau \sim 40$ minutes 
and the beam polarization might be different for each bunch.  
Therefore, the luminosity must be measured 
accurately for $\underline{\rm{each}}$ bunch over time intervals 
much less than $\tau$. In addition, a fast luminosity measurement is
essential for obtaining collisions, and monitoring and optimizing luminosity
during a fill.

The required accuracy in the luminosity measurement for HERA-II is $\sim 2\%$, 
which is similar to the most accurate luminosity measurement achieved at 
HERA-I.  The luminosity spectrometer described in this paper utilizes a new 
method of measuring luminosity at ZEUS which addresses the new problems 
of synchrotron flux, pile-up, and other requirements, 
while meeting the specifications for luminosity accuracy 
required by ZEUS physics goals.

In the spectrometer system, the bremsstrahlung photons are detected 
through their well understood pair conversion, 
$\gamma \rightarrow e^{+} + e^{-}$, in material of a beam-pipe exit window 
well downstream of the interaction region where these photons 
have been spatially separated from the circulating beams.  
After the converted electron pair 
\footnote{In this paper, ''electron'' is meant to specify both electron or 
positron.  Note that, aside from the direction of bend in the field, 
their interactions are identical at these energies.} 
has been spatially split by the magnetic field of a dipole magnet, 
the particles are individually detected by two small electromagnetic 
calorimeters 
placed at transverse distances separated from the direct synchrotron radiation
and unconverted bremsstrahlung beams. The observed rate of converted photons 
is proportional to the luminosity, as described in 
Sec.~\ref{sec:luminosity_calculation}.  

This setup reduces dramatically the requirements on data rate for the 
spectrometer 
calorimeters, since the primary photon beam bypasses them and the magnet 
insures that the large flux of low energy electrons from synchrotron photon 
conversions in material upstream of the magnet are swept away from the 
detectors.  Due to the small conversion fraction ($\sim 10\%$) in 
the window and the limited accepted energy range, the pile-up problem is 
reduced by two orders of magnitude, so that multiple observed photons 
constitute only a few percent of the rate even at the highest 
luminosities expected for HERA-II.  

With the relaxed rate requirements on the spectrometer calorimeters, 
their design can be sophisticated enough to simultaneously provide 
relatively precise measurements of the converted electron energies and 
positions
 --- in turn giving similarly precise information for the bremsstrahlung 
photon.  The redundancy and precision of the measurements provide important 
assurance on many aspects of the acceptance which must be known accurately 
with this technique.  In general, the technique is simple and accurate, 
and works well at high luminosities.

The structure of this paper is as follows: in 
Sec.~\ref{sec:spectrometer_design_and_components} are presented the details 
of the design of the luminosity spectrometer, including the calorimeter 
modules and a brief description of the data acquisition system.  
The detector calibration in a momentum analyzed electron test beam, 
and the reconstruction of photon properties from the calorimeter measurements 
are described in 
Sec.~\ref{sec:calorimeter_reconstruction_of_electrons_and_photon}.  
The method of calculating the luminosity is described in 
Sec.~\ref{sec:luminosity_calculation}, and the $in$ $situ$ calibration and 
operating performance during the first months of HERA-II operation are 
presented in Sec.~\ref{sec:system_operation}. 

\section{Spectrometer design and components}
\label{sec:spectrometer_design_and_components}
\begin{figure}[ht]
\begin{center}
\includegraphics[width=12.0cm]{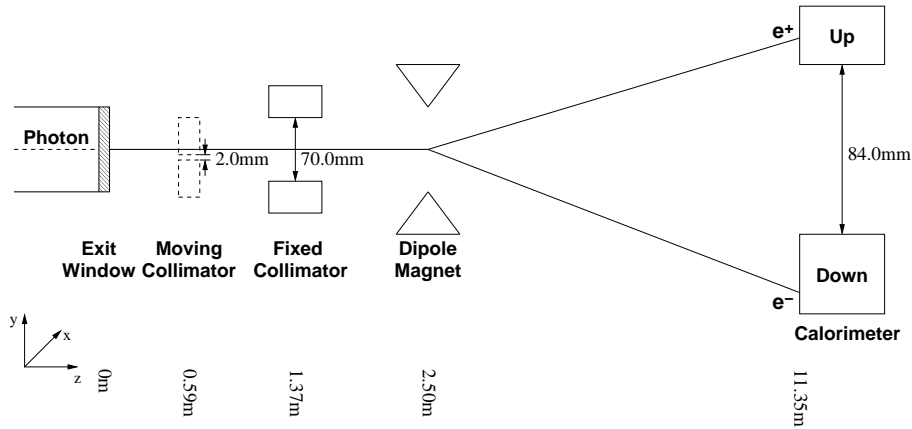}
\caption{Schematic showing major elements of the luminosity spectrometer. 
Note the very different scales for transverse and beam directions, 
and note that the exit window is 92.5~m downstream of the $ep$ nominal 
interaction point. The origin for the
transverse coordinates ($x$, $y$) was chosen at the center of the exit window.}
\label{fig:2-1-layout}
\end{center}
\end{figure}

A schematic layout of the spectrometer system is shown in 
Fig.~\ref{fig:2-1-layout}. The coordinate system is defined with the positive 
$z$-axis along the photon beam direction, $y$ pointing upwards and $x$ along 
the 
line away from the center of the HERA ring. Here we describe each component 
in the order encountered by a photon, beginning at the exit window of the 
photon beam pipe, through the collimators, magnet, and the calorimeters.  
Relevant aspects of the data acquisition system are also described.

\begin{figure}[ht]
\begin{center}
\includegraphics[width=8.0cm]{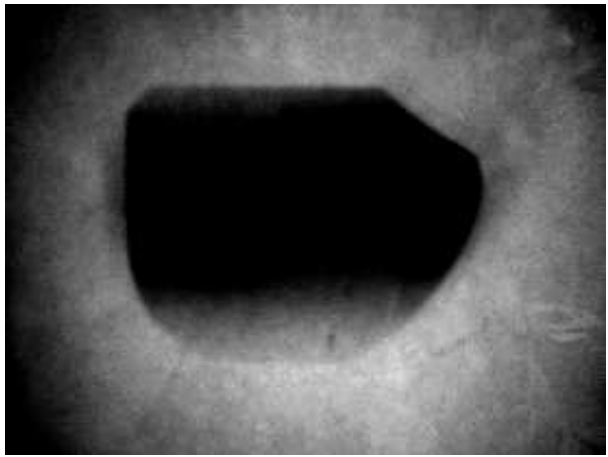}
\caption{Foil irradiated by synchrotron radiation. It was located near the 
front face of the calorimeters. 
The effects of aperture restrictions between the IP and the foil 
are clearly visible.}
\label{fig:2-2-aperture}
\end{center}
\end{figure}

The bremsstrahlung photons from the intersection region 
travel through a long 
vacuum pipe terminating at an exit window, at 92~m from the nominal 
intersection point (IP).  Upstream of the window inside the vacuum chamber 
(not shown in Fig.~\ref{fig:2-1-layout}) are various aperture 
restrictions such as collimators, magnets, and other apparatus.  
Figure~\ref{fig:2-2-aperture} shows the picture of a foil sensitive to 
synchrotron radiation, which was located in the beamline near the 
front face of the calorimeters.
The foil clearly indicates the 
aperture delimited by upstream obstacles.  For scale, the maximum horizontal 
extent of the aperture is 9~cm.  The maximum vertical limits are 
representative of the 7~cm aperture of the fixed collimator.

The window, at the upstream end of the spectrometer system, terminates the 
vacuum so that the remainder of the detection system is in air.  
Approximately $10\%$ of the 
photons convert into $e^{+}e^{-}$ pairs in the window.  
The typical energy of photons of 
interest is about $20$~GeV.  The distributions in $x$ and $y$ photon impact 
positions at the window within the acceptance region reflect primarily the 
angular divergence of the primary beam electrons as they collide in the 
upstream ZEUS interaction region.  

The converted electron pairs that successfully traverse subsequent collimators 
closely follow the original photon direction, until they encounter the 
magnetic dipole where they are split vertically.  
The magnetic field is such that 
a typical 10~GeV electron (positron) will then travel in the $\pm y$ 
directions with a vertical angle of about $10^{-2}$ rad, and encounter the 
calorimeters, whose vertical midpoints are displaced approximately $\pm 10$~cm 
from the incident photon beam.  The calorimeter modules permit measurements of 
the electrons' $x$ and $y$ 
positions with resolution of about 1.0 mm, and energies with resolution of 
about  $5\%$ 
at 10~GeV.  Calibrations in an electron test beam (described in 
Sec.~\ref{sec:calorimeter_reconstruction_of_electrons_and_photon}) and in 
operation with the spectrometer (described in 
Sec.~\ref{sec:system_operation}) verify these calorimeter measurement 
precisions.  

Events with electrons detected in both calorimeter modules are used to 
reconstruct the original photon's energy and transverse coordinates.  
The measured coincidence rate permits calculation of luminosity, 
as described in Sec.~\ref{sec:luminosity_calculation}.  The bremsstrahlung 
position distributions, while relevant to luminosity calculations, 
also provide a diagnostic of the interaction region for HERA operation.  

The locations of the different spectrometer elements 
were obtained from optical survey and verified with beam data. 
Transverse distances are known to a precision of order 1 mm 
while the dimensions 
along the direction of the beam are known with a precision of better 
than 1~cm.  These accuracies satisfy the requirements for the luminosity 
measurement.

\subsection{Exit window}
The composition of the exit window is
shown in Tab.~1:
\begin{table}[ht]
\begin{center}
\begin{tabular}{|c|c|}
\hline
\textbf{Material} & \textbf{Percent by Weight} \\
\hline
Aluminum & 85.3 \\
\hline
Silicon & 10.9 \\
\hline
Iron & 0.3 \\
\hline
Copper & 2.7 \\
\hline
Magnesium & 0.3 \\
\hline
\end{tabular}
\label{tab:exitwindow}
\caption{Components of the exit window, as measured by mass spectroscopy.}
\end{center}
\end{table}

The diameter of the window is 100 mm.  The window thickness was measured 
at several positions; the average thickness was $9.883 \pm 0.003$~mm 
with an RMS of 0.0054 mm, implying at least that level of uniformity.  
The fraction of converted photons in the window 
is therefore also uniform over the surface of the window. The variation in 
the cross section for photon energies 
between 5~GeV and 20~GeV is $\approx 1\%$ 
\cite{nist:xcom}; within the narrower energy window used for luminosity 
measurements, the energy dependence is 
very small and well simulated.

The radiation length of the exit window alloy is $X_{0}=8.23$~cm, which means 
that the exit window itself represents $0.12X_{0}$, so that about $8.8\%$ 
of interesting photons convert in the window.  Converted electrons traveling 
through the material may subsequently radiate and lose a fraction of their 
energy. This effect is small, well understood, and simulated accurately.
Also simulated is the multiple scattering 
of converted electrons, which leads to typical RMS deflections of  
$\theta_{x} \approx 2.8 \times 10^{-4}$ rad and a transverse spread at the 
calorimeter front surface of $\approx 3.4$ mm.  The resolution of reconstructed
photon conversion positions is dominated by this multiple scattering of 
electrons in the window.

\subsection{Collimators}
The collimators are used to select observed photons within a rectangular 
region inside the transverse dimensions of the vacuum pipe and within the 
uniform field region of the magnetic field.

The fixed collimator is a block of stainless steel, 30~cm long, with internal 
aperture $\Delta x \Delta y = 96 \times 70 \rm{mm}^{2}$.  It shields the 
calorimeters and the magnet from exposure to direct bremsstrahlung or 
synchrotron radiation photons from the circulating electron beam.  
However, measurements from synchrotron radiation indicate that beam elements 
upstream of the exit window more stringently limit the photon aperture, 
as shown in Fig.~\ref{fig:2-2-aperture} and verified with 
bremsstrahlung measurements described in Sec.~\ref{sec:luminosity_calculation}.

The moving (slit) collimator is a 15~cm long, water-cooled block of copper.  
When inserted, it restricts bremsstrahlung photons to pass through a narrow 
($\Delta x \Delta y = 110 \times 2 \rm{mm}^{2}$) slit
\footnote{The height was optimized for high rate while maintaining 
small edge effects.}.  
During normal running conditions, the moving collimator is out of the 
beam. 
When inserted it allows precise calibration of the
absolute energy scale of the calorimeter channels 
(see Sec.~\ref{sec:system_operation}).

\subsection{Dipole magnet}
The spectrometer uses a 60~cm long dipole magnet with a 10~cm horizontal 
aperture and typical field strength of $B_{x} \approx 0.5$ T.  
The bottom of the proton beam pipe passes only a few centimeters above the 
top of the bremsstrahlung aperture and the effect of the dipole field 
on proton beam operation must be 
minimal.  Hence magnetic shielding is arranged 
in the vicinity of the magnet around the proton beam-pipe.  
The shielding is a cylinder of diameter 85 mm, and centered at 
$x=17$ mm, $y=126.7$ mm with respect to our coordinate system (and 
the center of the magnet).
\begin{figure}[ht]
\begin{center}
\includegraphics[width=8.0cm]{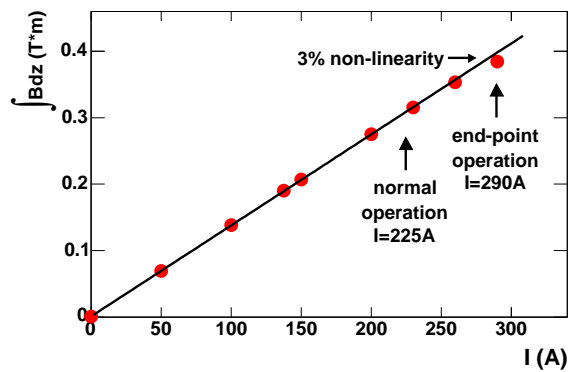}
\caption{The points show data of the magnetic field integrated along the beam 
direction as a function of the dipole excitation current. The magnet is very 
linear in the region of normal operation, and only slightly non-linear at the 
highest currents, where measurements of the bremsstrahlung end point were made.
}
\label{fig:2-3-magnetam}
\end{center}
\end{figure}

\begin{figure}[ht]
\begin{center}
\includegraphics[width=8.0cm]{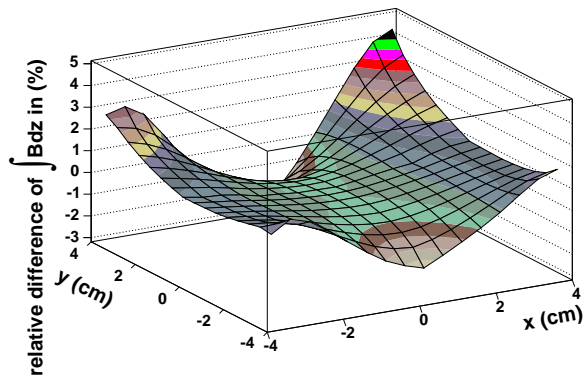}
\caption{The difference between the measured integrated magnetic field at 
nominal dipole operating current, and the integrated field fitted by the 
OPERA program as a function of transverse positions.}
\label{fig:2-3-magnetbm}
\end{center}
\end{figure}

The magnetic field was measured, using a Hall probe, as a function of position 
at several different magnet currents \cite{misc:holler:private}.  
The integrated magnetic field in the 
central region versus excitation current, 
shown in Fig.~\ref{fig:2-3-magnetam}, 
indicates that the field is linear over the range of normal operation,
and saturates slightly at the highest excitation currents used.  
The position dependent measurements have been interpolated using 
Maxwell's equations supplemented with a Jacobi relaxation 
method to obtain the full and continuous magnetic field map.
This map was implemented in the simulation program.
The systematic errors have been estimated using the OPERA program 
\cite{opera,misc:marx:private}; 
the differences are shown in Fig.~\ref{fig:2-3-magnetbm} 
over a region 
which encompasses the full transverse aperture.  Uncertainties thus estimated 
were typically less than $1\%$ throughout the acceptance region encompassed 
within $96 \times 70$ mm, with an average uncertainty less than $0.5\%$ 
near the region of highest flux.  Therefore, the magnetic field is 
well understood and creates an even smaller uncertainty on the precision of 
the luminosity measurement.

\subsection{Calorimeters}
The two calorimeter modules were originally constructed as 
``Beam pipe calorimeters'' (BPCs) in ZEUS to measure DIS processes at very 
low values of $Q^{2}$ \cite{Breitweg:2000yn}.  They were designed to measure 
positions and energies of electrons near the HERA beam energy (27.5~GeV).  
The original description of their construction can be found 
in \cite{thesis:surrow:1998,thesis:monteiro:1998}.  For this application, one 
module was rebuilt in order to fit within the space constraints. 
The main parameters 
are listed in the Tab.~\ref{tab:calorimeter} and the relevant properties 
and changes are elaborated here.
\begin{table}[ht]
\begin{center}
\begin{tabular}{|c|c|}
\hline
\multicolumn{2}{|c|}{\textbf{Calorimeter Specifications}}\\
\hline
Depth & $24 X_{0}$ \\
\hline
Moliere radius & 13mm \\
\hline
Energy resolution (stochastic term) & $17\% \sqrt{E}$ \\
\hline
Energy scale calibration precision & $0.5\%$ \\
\hline
Energy uniformity & $0.5\%$ \\
\hline
Linearity & $\leq 1\%$ \\ 
\hline
Position resolution & $< 1$ mm \\
\hline
\end{tabular}
\caption{Physical characteristics of the calorimeter modules.}
\label{tab:calorimeter}
\end{center}
\end{table}

Both BPC modules are segmented tungsten-scintillator sampling calorimeters, 
permitting simultaneous measurement of the electron's energy and transverse 
position of impact.  The passive layers contain 26 plates of 3.5 mm thick 
tungsten alloy.  These 24 electromagnetic radiation lengths ($X_{0}$) are
 more than 
adequate for longitudinal containment in this application.  The active 
elements consist of scintillator fingers, alternating after each plate 
in the $x$ and $y$ directions, each finger 7.9 mm wide and 2.6 mm thick.  
One end of each finger is aluminized to provide an efficient end reflector. 
Each scintillator is optically decoupled from its neighbors by a wrapping 
of $27.5$ $\mu\rm{m}$ aluminum foil.  The net effective width of each finger, 
including wrapping and air gaps, is 8.0 mm.

The other end of each scintillator finger rests against a wavelength 
shifting (WLS) bar 7 mm wide and 2 mm thick.  The scintillator fingers 
oriented in alternate layers, in the longitudinal direction, are observed 
with one WLS bar representing a single readout channel.  Each bar is viewed 
by a photomultiplier (PMT) from the rear. The WLS is aluminized at the front 
end to reduce attenuation effects.  To compensate for remaining attenuation 
in the WLS bars, they were wrapped in reflective correction masks to ensure 
longitudinally 
uniform response from each scintillator located along the length of a bar. 
The geometry of some WLS bars required redesign for this application; 
they were bent and pointed toward the new locations of the 
Hamamatsu R5600U-03 PMTs \cite{Hamamatsu}.

Each module has 16 channels for $x$-position reconstruction. For the 
$y$-position, 
the lower module has 15 channels and the upper 11 channels, reflecting 
limitations imposed by the small separation between the photon beam and the 
proton beam pipe.  The 84 mm transverse separation of the modules is 
maintained by rigid brass bars. Since the calorimeter containment 
vessels incorporate 
5 mm of tungsten on surfaces adjacent to the beam, the net transverse 
distance between inner scintillators of the two modules is 94 mm.  
These positions and the trajectories of electrons after traversal of the 
magnetic field are discussed more fully in 
Sec.~\ref{sec:calorimeter_reconstruction_of_electrons_and_photon}.  

In order to protect the calorimeters from radiation due to scattered 
synchrotron radiation, the unconstrained sides and rear were shielded by 
bricks and sheets made of lead.  Also, a thin (5 mm) sheet of 
lead was placed 
in front of the modules. A system of light emitting diodes (LEDs) 
provides fast light pulses to all PMTs of a module simultaneously under 
remote command.  This system permits the relative PMT gains to be monitored 
for short term changes.  The LED system also 
provided an independent monitor of deadtime effects.  

The relative channel-to-channel gains were initially 
set using test beam electrons, described in 
Sec.~\ref{sec:calorimeter_reconstruction_of_electrons_and_photon}.  
The relative response of each channel was monitored $in$ $situ$ with 
calibration runs as described in Sec.~\ref{sec:system_operation}.  
The calibration runs were important, 
especially during early running when unstable beam operation created 
changes in channel-to-channel response.  (These were found to primarily arise 
from irradiation of WLSs near the PMTs.)  Changes were monitored and 
compensated to maintain the relative gains of all readout channels within a 
module as well as the relative average gains of the two modules.  

\subsection{Data Acquisition System}
\label{sec:daq}
One important challenge in this application is accurate digitization of the 
calorimeter PMT signals, which are separated by the 96~ns HERA bunch spacing.
In order to minimize noise, analog processors were located in the HERA tunnel 
close to the detectors.  These electronics are inaccessible during normal HERA 
running.  The digital electronics is accessible, located in the ZEUS main hall 
about 100~m upstream of the luminosity detectors. The spectrometer  
data acquisition system (DAQ) was designed and constructed by the ZEUS 
Krakow group \cite{misc:wojtek:private} and DESY electronics 
development group,
and is common to the luminosity calorimeter 
system and the 6 meter tagger\cite{misc:calor2000:6mtag}.  
The latter system records the energies of final state bremsstrahlung electrons 
with energies complementary to those accepted by the spectrometer.  

\begin{figure}[ht]
\begin{center}
\includegraphics[width=12.0cm]{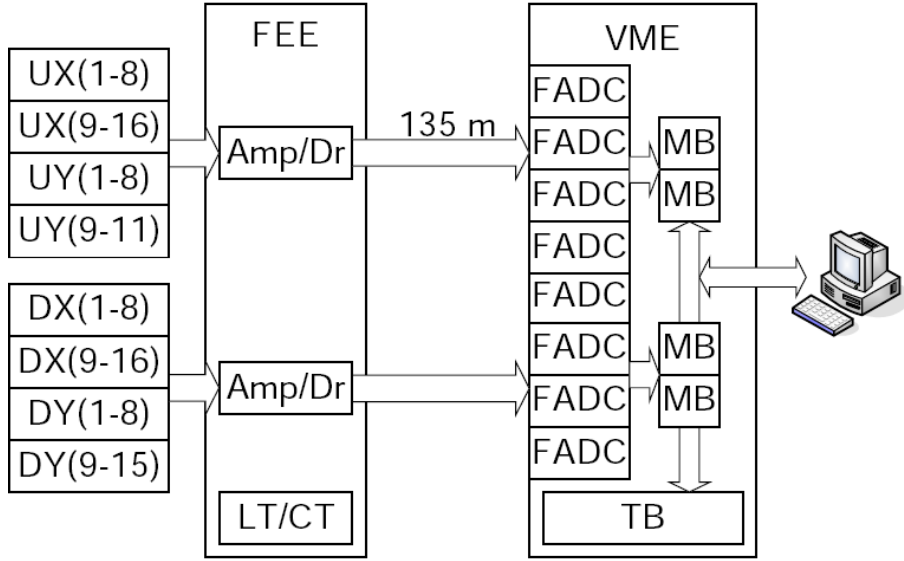}
\caption{Schematic of the luminosity spectrometer DAQ system. The calorimeter 
signals are sent to the nearby FEE system. The analog signals then travel to 
the ZEUS hall to be processed by the digital system in a VME crate.}
\label{fig:2-5-daq}
\end{center}
\end{figure}

Figure~\ref{fig:2-5-daq} outlines the full system.  The digitized signal is 
used immediately in the triggering system and retained for the luminosity 
calculation.  In more detail, the readout system consists of the following 
components:
\begin{itemize}
\item
The front-end electronics (FEE) is the analog system of amplifiers and 
drivers used to shape and transmit the analog PMT signals to the 
digital system.
\item
The digital electronics includes eight flash analog-to-digital 
converter (FADC) 
boards, four memory boards (MB), and one trigger board (TB).  Each MB performs 
operations on the data as well as storage.  All are located in a single 
VME crate.  
\item
A PC, operating under LINUX, for online processing of the data and calculation 
of the ZEUS luminosity.
\end{itemize}

The signal from each PMT is transferred through a 6~m long 
cable to the FEE crates. 
The FEE 
amplify and shape these signals and drive them through the 
135~m long RG213 transmission lines to the digitization units in the ZEUS hall.
Additionally, the FEE crates each contain driver modules for an LED 
(input to the PMT cathodes) and charge-injector to the FEE amplifiers.   
These can be independently triggered to test both the PMTs and the remainder 
of the DAQ system.  The analog signals are integrated with a time constant of 
65~ns and subsequently digitized by 12-bit FADC modules operating at about 
10 MHz, synchronously with the HERA beam bunches. Each FADC board consists 
of a pole-zero filter, which shapes the input analog signal to reduce the 
decay tail. Digitized values from each PMT for each bunch 
crossing are transferred 
to the 16-channel MBs residing in the same VME crate.  
The MB performs the following operations on each signal:
\begin{itemize}
\item
A subtraction is made of the stored digital value from the previous bunch 
crossing to provide a dynamic pedestal subtraction.  
In the process, low frequency noise (compared to 10 MHz) is removed. 
The low rates in all channels of the spectrometer ensure few mistakes 
due to signals in consecutive bunches.
\item
Sums are made of all the channels in each dimension of 
the upper ($x_{up}$, $y_{up}$) and lower 
($x_{dn}$, $y_{dn}$) calorimeters, resulting in four energy 
sums ($E_{x}^{up}$, $E_{y}^{up}$, $E_{x}^{dn}$, $E_{y}^{dn}$) 
in each crossing contained in the four MBs.  
These sums are retained with the signals from the individual channels.  
The sums are also sent to the TB where they are used for 
triggering decisions (data retention) on each beam crossing.
\item
Meanwhile the data have been placed in a buffer to await the trigger decision.
For events in which the TB flags the event as useful, the data are copied 
sequentially to output buffers.  These buffers are read continuously 
via VME at a rate up to 10 kHz.
\end{itemize}

The TB, in the VME crate with the MBs, communicates with the MBs through a 
custom backplane.  The programmable TB decides when an event satisfies 
specified criteria; if so, it sets an ACCEPT signal on the backplane.  
The criteria utilize two digital sums: one ($E_{up}$) from all the $x_{up}$ 
and $y_{up}$ 
channels; the other from the corresponding sum ($E_{dn}$) 
in the down detector.  
Only events that pass a nominal threshold value in each of the two detectors 
are passed on for further consideration.

If the MB buffer were to become full, data acquisition must halt until buffer 
space is available.  This is a situation requiring deadtime corrections, 
which are best minimized.  Therefore, at the highest luminosities, a 
trigger prescale 
factor is applied to maintain an average rate well below the maximum trigger 
rate of 10 kHz.  The TB provides counters that record (1) the total number of 
HERA bunches and (2) the number in which the MB buffer is not full and active.
When the MB buffer is full and incoming data must be discarded, the 
active-time counter stops counting. These counter values, sent to the online 
PC each second, provide a continuous deadtime correction for the luminosity 
calculation and also permit this deadtime to be monitored.  
The deadtime correction and the prescale factor are independently measured 
using light test events by recording both the LED signal rate and the rate 
at which these events are recorded.  
The prescale factor was set to 3 during HERA high luminosity running.

Data from accepted events are transferred to output MB buffers and sent 
via VME bus to the online PC.  The information is used to reconstruct the 
energy and position information for the detected photon, as described in 
Sec~\ref{sec:calorimeter_reconstruction_of_electrons_and_photon}, and is 
stored in histograms for monitoring purposes.  The ZEUS luminosity is 
calculated offline using histograms and counters saved to tape every 
16 seconds, as described in Sec.~\ref{sec:luminosity_calculation}.

\section{Calorimeter Reconstruction of Electrons and Photons}
\label{sec:calorimeter_reconstruction_of_electrons_and_photon}

\subsection{Reconstructing electron energies and positions}
The functions of the calorimeter modules are to measure the energies and 
transverse coordinates of the two electrons, respectively in the up ($up$) and 
down ($dn$) calorimeters, using the information recorded from the 
PMT channels associated with the horizontal and vertical 
strips. The calorimeters were extensively tested in a DESY-II electron 
beam at energies up to 6~GeV, prior to installation in the HERA tunnel.  
Described here are the algorithms used to transform the charges recorded by 
the PMTs into 
the incident electron energy (energy calibration) and the transverse position 
coordinates, as well as the tests performed in the electron test beam.  
The measurements show that the calibration precision is more than adequate for 
the present task.  The position and energy resolutions, leakage 
effects, as well as remaining attenuation in the scintillator strips were 
also measured with the test beam. The measured leakage and attenuation effects
with test beam have been used to correct the energy reconstruction of data. 
Further calibration checks on the 
modules were performed in the spectrometer configuration, as described in 
Sec.~\ref{sec:system_operation}, in order to maintain the calibration during 
operation.

The preliminary settings of the PMT gains were set using an LED system.  
Also, by varying the high voltages applied to the PMTs, an empirical function 
was obtained providing the relative gain variation 
with high voltage, independently for 
each channel.  Additional tests, including absolute calibrations, 
were carried out in the test beam.

The test beam setup located each module on a movable table, which could be 
translated horizontally and vertically by known amounts with a stepping motor.
The electron test beam was focused, steered, and momentum defined by magnetic 
beam elements. Counters constrained the beam to be within a transverse circle 
of 3.0 mm diameter. Initially, the module was located with the 5.0~GeV beam 
incident near its center. The gains of the PMTs were set to nominal values, 
and the resulting net charge (in ADC counts) from each PMT channel was 
recorded for each event as the module was scanned horizontally 
or vertically.  Using these scans, the PMT gains were then adjusted, 
by changing the applied voltage according to the empirical functions 
previously measured, so that the peak charge was the same for all channels.
This procedure converged after a single iteration.  

\begin{figure}[ht]
\begin{center}
\includegraphics[width=8.0cm]{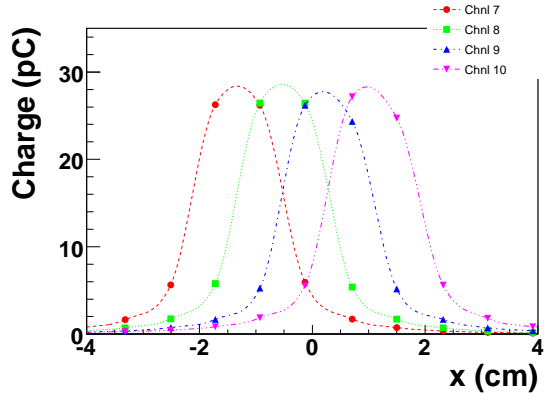}
\caption{Responses of four adjacent calorimeter channels while the beam was 
scanned across the $x$-coordinate. Each channel is represented by a unique 
symbol.}
\label{fig:3-1-btest_x_scan}
\end{center}
\end{figure}

Figure~\ref{fig:3-1-btest_x_scan} shows typical data (points) from four 
adjacent channels in such a horizontal scan.  
The points joined by a continuous curve represent a single channel's 
response.  Recalling that an individual scintillator width is 8.0 mm, 
the figure also illustrates that  electron showers typically deposit a 
substantial fraction of their energy in at least two channels, but very 
little energy is outside four channels.  

By repeating such horizontal scans at different vertical positions, 
any uncompensated light attenuation in the scintillator was measured so its 
effects could be corrected during analysis.  The overall gains were set so 
that the total energies for electrons incident at the midpoint of the module 
from the horizontal and vertical channels were equal. From the known beam 
energy, the charge response was calibrated directly to the sampled energy 
as a parameter $S_{i}$ for the $i^{th}$ channel as $S_{i}=$ energy divided 
by average recorded charge.  These parameters were then used to reconstruct 
shower energy as described below.

To minimize noise, only strips near the shower maximum were actually used 
for reconstruction.  The energy of the incident electron sampled from 
the $x$-strip sum is 
\begin{equation}
E_{x}=C_{trans}^{x}(x_{e},y_{e}) C_{att}^{x}(y_{e}) 
\sum_{x-strips}^{4} \varepsilon_{i}^{x} S_{i},
 \label{eqno1}
\end{equation}		
where the sum runs over a cluster of the four highest energy adjacent strips;  
$\varepsilon_{i}^{x}$ is the charge from the $i^{th}$ channel (in ADC counts),
$C_{trans}^{x}(x_{e},y_{e})$ corrects for transverse energy leakage 
(a small effect except at the edges of the calorimeter) 
and $C_{att}^{x}(y_{e})$ corrects 
for the small residual attenuation as measured in the test beam. 
The positions ($x_{e},y_{e}$) of the shower centroid were also obtained from 
the event, as described below.  The measurement of $E_{y}$, the energy 
sample of the $y$-strips, was obtained in a similar manner. 

The total energy of the electron was
\begin{equation}
E_{e}=E_{x} + E_{y}.
 \label{eqno2}
\end{equation}		
The position coordinates were calculated using linear energy-weighted means 
over the deposited energy. For the upper calorimeter, 
the internal $x$-coordinate for the shower, $x_{e}^{up}$, was obtained from 
\footnote{A logarithm weighted algorithm though slightly more precise was not 
used here because it was
found to be too time consuming for the online processing, and it is not
well described by the simulation.}
\begin{equation}
x_{e}^{up}=\frac { \sum_{x-strips} X_{i} \varepsilon_{i}^{x} S_{i}} 
{\sum_{x-strips} \varepsilon_{i}^{x} S_{i}},
 \label{eqno3}
\end{equation}		
where $X_{i}$ is the known central location of the $i^{th}$ channel strip 
within the calorimeter.  Only channels above a threshold energy of 60 MeV 
were used for these sums.  The $x$ position in the spectrometer coordinate 
system for the electron in the up detector, is given by 
\begin{equation}
x_{up}=x_{e}^{up}+\Delta x^{up},
 \label{eqno4}
\end{equation}		
where $\Delta x^{up}$ is the known alignment offset relative to the coordinate 
system of Fig.~\ref{fig:2-1-layout} for this calorimeter.  
The shower $y$-coordinate, $y_{up}$, was obtained similarly.

\begin{figure}[ht]
\begin{center}
\includegraphics[width=8.0cm]{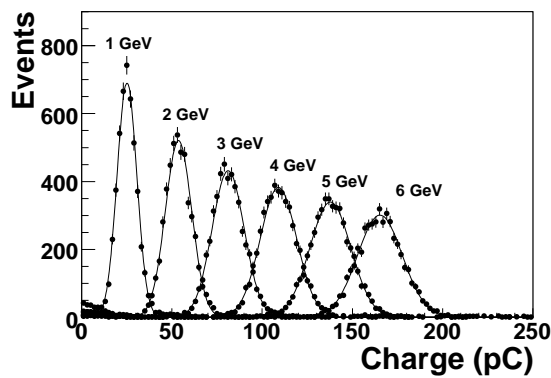}
\caption{Spectra of total reconstructed PMT charge in the calorimeter module 
for six different energy settings of the test beam.}
\label{fig:3-2-e-resp}
\end{center}
\end{figure}
\begin{figure}[ht]
\begin{center}
\includegraphics[width=8.0cm]{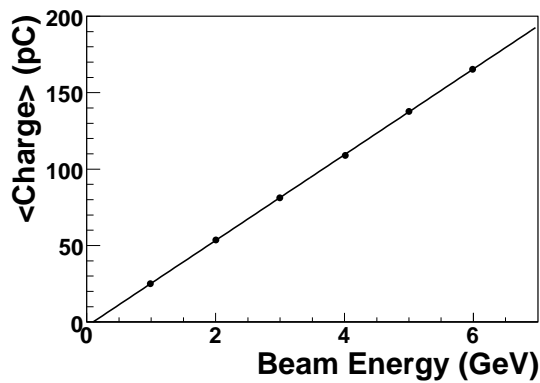}
\caption{The mean total charges (from the spectra of Fig.~\ref{fig:3-2-e-resp})
as a function of selected beam energy. The response is linear.}
\label{fig:3-3-e-linear}
\end{center}
\end{figure}

The reconstructed energy response was measured with differing known test 
beam energies; the spectra for six different energies are shown in 
Fig.~\ref{fig:3-2-e-resp} and illustrate that the measurements show the 
expected results.  Figure~\ref{fig:3-3-e-linear}, 
showing the mean response as a function of the beam energy, 
demonstrates that the response is linear up to at least 6~GeV.  
The intrinsic linearity of calorimetry and the tested linearity of the 
electronics implies linearity for even higher energies.

\begin{figure}[ht]
\begin{center}
\includegraphics[width=8.0cm]{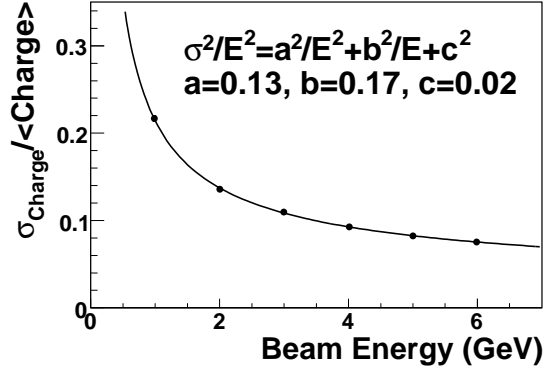}
\caption{The RMS spreads of the spectra in Fig.~\ref{fig:3-2-e-resp} as a 
function of beam energy. The results of a fit to the displayed function
are also shown.}
\label{fig:3-4-e-resol}
\end{center}
\end{figure}

Figure~\ref{fig:3-4-e-resol} depicts the RMS deviation from the mean as a 
function of energy: the parameterization indicates that in the region of 
intended measurement, around 9~GeV, the RMS is mainly proportional to 
the stochastic $\sqrt{E}$ term, as expected for a sampling calorimeter.

\subsection{Reconstructing bremsstrahlung energy and position}
\label{sec:recon_photon}
The operation of the calorimeter modules in the bremsstrahlung beam is 
discussed in detail in Sec.~\ref{sec:system_operation}.  We note that, 
in this geometry where the $up$ and $dn$ calorimeters measure the two 
electrons, the original photon may be directly reconstructed from 
these measurements. The energy of the photon is
\begin{equation}
E_{\gamma}=E_{up} + E_{dn}.
 \label{eqno5}
\end{equation}		
The transverse $x$-position of the photon is
\begin{equation}
x_{\gamma}=\frac{1}{2} [x_{up} + x_{dn}]
 \label{eqno6}
\end{equation}		
and the transverse $y$-position (bend plane of magnet) is 
\begin{equation}
y_{\gamma}=\frac{E_{up}y_{up} + E_{dn}y_{dn}}{E_{up} + E_{dn}}.
 \label{eqno7}
\end{equation}		
The energy weighting in the last equation arises because the magnet imparts 
equal transverse momentum to each electron.

\section{Luminosity Calculation}
\label{sec:luminosity_calculation}
Colliding beam luminosity is quantitatively related to the circulating beam 
currents and the spatial overlap as they intersect.  By definition, the 
integrated luminosity in the collision region provides the normalization for 
any cross section so that, if $N_{int}$ events of a specific process are 
observed during the time period during which the integrated luminosity is 
known, then the cross section for that process is
\begin{equation}
\sigma = N_{int} / L_{int},
 \label{eqno8}
\end{equation}		
where $L_{int}$ is typically specified in units of inverse picobarns 
$\rm{pb}^{-1}$.

\subsection{Instantaneous and specific luminosities}
The instantaneous luminosity ($L_{inst} = dL_{int} / dt$) is defined, 
for the purposes of this paper, to be the luminosity averaged over the 
period between accelerator bunch crossings (96~ns).  
Since $L_{inst}$ scales with the product of the electron ($I_{e}$) and 
proton ($I_{p}$) currents, it is convenient to define the specific 
instantaneous luminosity 
\begin{equation}
L_{spec} = \frac{L_{inst}} {\sum_{i} I_{e}^{i} I_{p}^{i}},
 \label{eqno9}
\end{equation}		
where the sum includes the currents in all colliding bunches.

\subsection{Luminosity Measurement}
\begin{figure}[ht]
\begin{center}
\includegraphics[width=8.0cm]{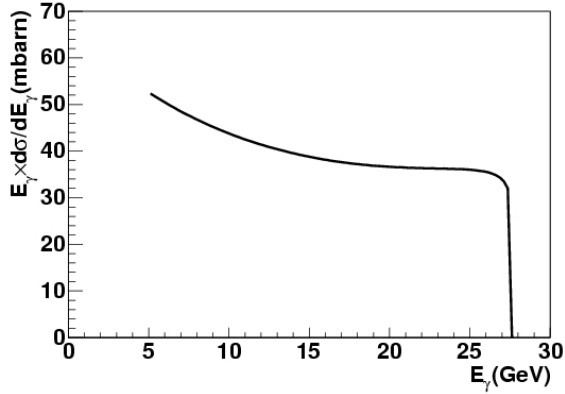}
\caption{The Born level Bethe-Heitler cross section multiplied by 
$E_{\gamma}$. The cutoff at the 
electron beam energy (27.5~GeV) is readily apparent.}
\label{fig:4-1-betaheitler-new}
\end{center}
\end{figure}
The most accurate method for measuring luminosity is to measure the rate 
from a process whose cross section is large and 
well known.  At HERA, luminosity is calculated 
from the measured rate, $R$, of photons created in the bremsstrahlung 
process $ep \rightarrow ep \gamma$.  The energy spectrum of bremsstrahlung 
photons is accurately described by the Bethe-Heitler 
formula \cite{prslon:a146:83}, 
shown in Fig.~\ref{fig:4-1-betaheitler-new} as the product of energy and 
differential cross section, in millibarns, as a function of photon energy.  
Additional corrections from radiative processes affecting the spectrum at 
energies above 5~GeV are well understood and incorporated into calculated 
cross sections.  The photons of interest here are those whose converted 
electrons have energies appropriate to enter and trigger both calorimeter 
modules.  As discussed below, the nominal magnetic field and geometry 
are such that the photon energies of interest are typically near 20~GeV.

Luminosity is obtained from the equation
\begin{equation}
L_{inst} = \frac{R} {f A \sigma},
 \label{eqno10}
\end{equation}		
where $\sigma$ is the bremsstrahlung cross section integrated over the 
same energy interval as the measurement of $R$.  The parameter, $f$ , 
represents the fraction of photons in the bremsstrahlung beam that convert 
into electron pairs in the material upstream of the magnet, primarily in the 
exit window.  The acceptance, $A$, includes the overall probability 
(not included in $f$) for a photon created at the intersection region to 
be observed as an $e^{+} e^{-}$ pair in the spectrometer.  
Imprecise knowledge of $A$ creates the largest uncertainty in the measurement, 
as will be discussed in detail below.  The acceptance of the luminosity 
system depends primarily on two issues:
\begin{itemize}
\item 
obstacles upstream of the entry to the luminosity system that remove photons 
from the edges of the bremsstrahlung beam; 
\item
the fraction of the converted pairs in which both electrons are accepted 
into the fiducial areas of the calorimeters.  
\end{itemize}
The conversion fraction and these two contributions to the acceptance 
will be discussed in turn.

\subsection{Conversion fraction, f}
\begin{figure}[ht]
\begin{center}
\includegraphics[width=8.0cm]{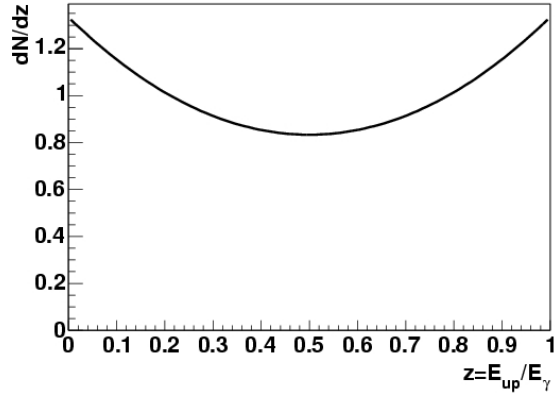}
\caption{The distribution of the fraction of the photon energy carried
by one of the electrons after conversion.
}
\label{fig:4-2-raw-z-dist}
\end{center}
\end{figure}
The fraction of photons converted into $e^{+} e^{-}$ pairs depends on 
(a) the absolute and energy-dependent pair production cross section and 
(b) the details of the conversion material in the exit window and the air 
between the exit window and the magnet.
\footnote{The gas in the entire vacuum pipe upstream of the exit window 
is negligible.}  
The materials and thickness of the exit window are well understood, 
and described in detail in Sec.~\ref{sec:spectrometer_design_and_components}.  
The total pair production cross section on each nucleus is largely independent 
of photon energy and is given to good accuracy by a simple formula  that 
depends on the radiation length and other well-understood 
properties of the converting material. The 
differential cross section for one electron of the pair to retain a fraction, 
$z$, of the photon energy, shown in Fig.~\ref{fig:4-2-raw-z-dist}, 
is also largely independent of energy.

The conversion efficiency and acceptance used in this analysis were 
calculated using the full GEANT 3.21 simulation.
Subsequent propagation and interactions of the electrons used the fully 
parameterized magnetic field map discussed in 
Sec.~\ref{sec:spectrometer_design_and_components}.   

\subsection{Upstream Obstacles}
\begin{figure}[ht]
\begin{center}
\includegraphics[width=8.0cm]{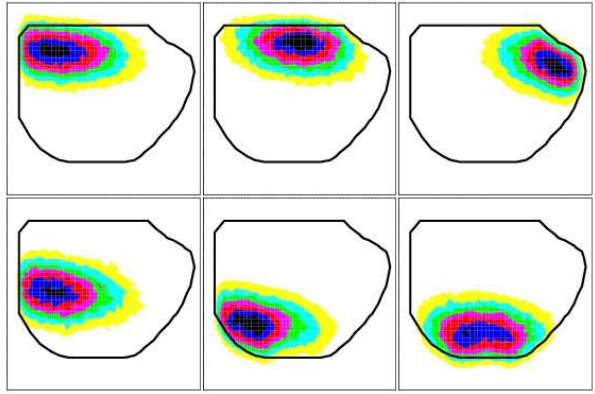}
\caption{Probability contours showing densities of observed events as the 
electron beam was steered to center the bremsstrahlung beam near the aperture 
edges. The data corroborate the aperture (solid contour) obtained using 
synchrotron radiation.}
\label{fig:4-3-tilt-scans}
\end{center}
\end{figure}

The acceptance includes a correction for incoming photons obscured by upstream 
obstacles.  The unobscured 
region was delineated in two independent ways: (a) the aperture obtained 
from a foil sensitive to synchrotron radiation described in 
Sec.~\ref{sec:spectrometer_design_and_components}; and (b) measurements of 
the transverse tails of the bremsstrahlung beam obtained by moving its  
centroid across the aperture. The two methods agree in detail as shown 
in Fig.~\ref{fig:4-3-tilt-scans}, where the sharp contour 
delineates the $x-y$ region 
indicated by illumination of the foil shown in 
Fig.~\ref{fig:2-2-aperture}.  The elliptical contours show regions 
observed from bremsstrahlung photons due to beam collisions at the 
intersection region as the electron beam was steered through various 
incident angles.  The regions from the bremsstrahlung runs in 
Fig.~\ref{fig:4-3-tilt-scans} can be seen to outline the same aperture as that 
indicated from synchrotron radiation.  

\begin{figure}[ht]
\begin{center}
\includegraphics[width=8.0cm]{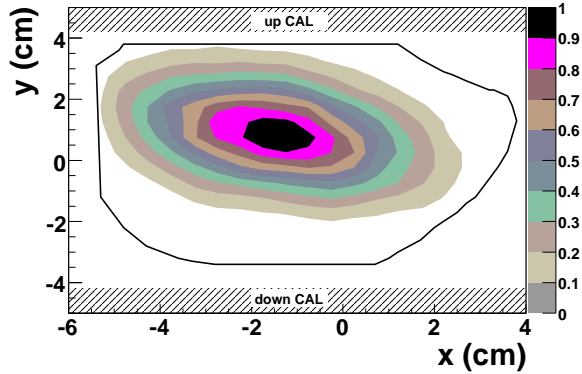}
\caption{Contours of equal probability for the bremsstrahlung beam in its 
nominal location. The contours represent relative probabilities as indicated 
by the key on the right. Note the aperture and vertical locations of 
calorimeter modules.}
\label{fig:4-4-contour}
\end{center}
\end{figure}

With beams steered in their nominal directions, Fig.~\ref{fig:4-4-contour} 
shows a typical 
measurement of the $x-y$ contours due to bremsstrahlung produced by the 
intersecting beams.  Note that the intersecting beam profile has a clear 
elliptical shape, wider in the horizontal than the vertical. 
The major axis 
of the ellipse is seen to be rotated around the beam direction. 
The detailed shape was found to fit 
orthogonal Gaussians at axes rotated by about 10 degrees to the horizontal.  
Typical distributions, projected along the orthogonal rotated axes, 
had standard deviations about 2~cm along the more horizontal axis and 1~cm 
along the more vertical axis.  The fitted 1D distributions are discussed 
further in Sec.~\ref{sec:system_operation}.

The acceptance correction for the photons blocked by the upstream aperture 
typically amounted to $5-10\%$, depending primarily on the photon beam 
properties at the time.  

\subsection{Converted Pairs Accepted into Calorimeter}
The spectrometer acceptance depends on the fraction of photons converted 
upstream of the magnet for which $\underline{\rm{both}}$ electrons of 
the pair enter the calorimeters.  This acceptance depends strongly on 
the distribution in $z$ shown in Fig.~\ref{fig:4-2-raw-z-dist} and on the 
energy of the interacting photons, $E_{\gamma}$, because of the deflection 
of each electron by the magnetic field. 

\begin{figure}[ht]
\begin{center}
\includegraphics[width=8.0cm]{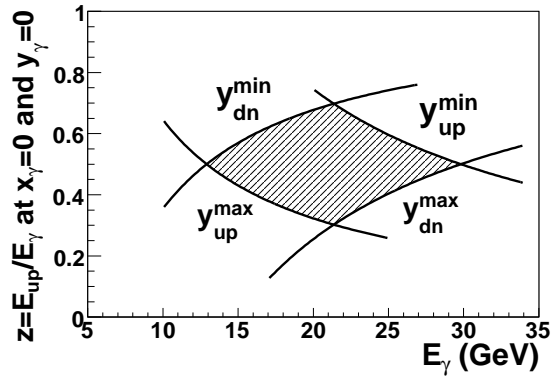}
\caption{Curves for Eqn.~\ref{eqno12} delineating the restrictions imposed by 
the range of accepted $y$-positions in the two calorimeters for an $e^{+}e^{-}$
conversion originating vertically equidistant from the calorimeter modules. 
Only the shaded region has an observed photon
within the fiducial region of both calorimeters which is therefore accepted.}
\label{fig:4-5-z-vs-egam-contours}
\end{center}
\end{figure}

Most of the acceptance dependence on $z$ and $E_{\gamma}$ can be understood 
from elementary arguments. Consider photons traveling along a beam axis
transversely located midway between the calorimeter modules
with $y=0$ \footnote{The actual vertical coordinate origin described here
is offset about 0.5 cm in our system. (See Figs.~\ref{fig:2-1-layout} and 
~\ref{fig:5-3-nup-ndn}.) 
A hypothetical simplified case is used here for pedagogical reasons.}.  
Each electron acquires the same transverse momentum, 
in opposite directions, from the field traversal: $p_{T}=0.3 \int B_{x}dz$, 
where the integral represents the integrated magnetic field (in Tm) 
along the path through the magnet.  At the nominal running current (225 A), 
the value of $p_{T} \approx 0.1$~GeV.  The deflection of an electron of 
momentum $p$ (in GeV/c) in the magnetic field can be approximated as 
\begin{equation}
y=\ell \frac{p_{T}} {p},
 \label{eqno11}
\end{equation}		
where $\ell$ is the distance from the center of the magnet to the calorimeters.
Only when the values of $y$ for each electron are such that both enter ``good''
calorimeter regions are events accepted.  If the fraction of the photon 
energy acquired by the upwardly deflected electron is $z=p/E_{\gamma}$, 
the other electron acquires a fraction $1-z$ of the photon energy.   
It follows that $z$ and $E_{\gamma}$  
are directly related to the electrons' locations. 
The locations of the electrons in the upper ($y_{up}$) and lower ($y_{dn}$) 
calorimeter modules are related to $z$ and $E_{\gamma}$ as follows:
\begin{equation}
zE_{\gamma}=\frac{\ell p_{T}}{y_{up}} \;\;\;\;\;\;\;\;\;\;\;\;\;\;\; 
(1-z)E_{\gamma} = \frac {\ell p_{T}}{y_{dn}}.
 \label{eqno12}
\end{equation}		
The inner and outer $y$-limits of the fiducial areas of the two calorimeters 
therefore delineate limits in the plane of $z$ versus $E_{\gamma}$ that are 
accepted.  This is illustrated in Fig.~\ref{fig:4-5-z-vs-egam-contours}, 
where the limits due to the 
outer and inner $y$-limits of the upper and lower calorimeter modules define 
the accepted region to be that internal to the curves. 

\begin{figure}[ht]
\begin{center}
\includegraphics[width=8.0cm]{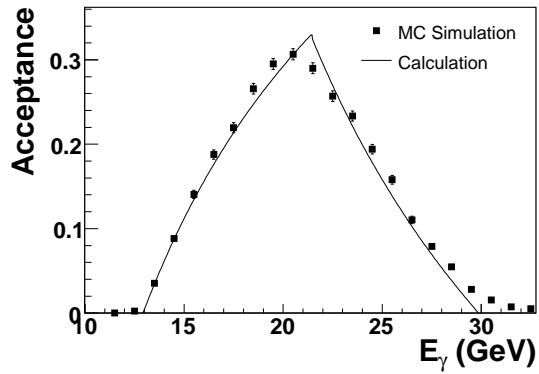}
\caption{The solid line represents the fraction of conversions shown in 
Fig.~\ref{fig:4-2-raw-z-dist} integrated over the accepted region of 
Fig.~\ref{fig:4-5-z-vs-egam-contours} for an $e^{+}e^{-}$ originating 
at $y=0$. The points result from the Monte Carlo predictions for the 
same case, which include actual conditions and responses.}
\label{fig:4-6-egam-acc-from-origin}
\end{center}
\end{figure}

If all photons arrived with coordinate $y_{\gamma}=0$, the acceptance at 
fixed $E_{\gamma}$ would be the integral of the function 
in Fig.~\ref{fig:4-2-raw-z-dist} over the accepted $z$-range shown 
cross-hatched 
in Fig.~\ref{fig:4-5-z-vs-egam-contours}.  
This simple acceptance function is shown as the solid line 
in Fig.~\ref{fig:4-6-egam-acc-from-origin}.  
The result of a more detailed Monte Carlo 
calculation for this beam configuration, the points in the same figure, 
shows that this simple model gives the major features of the acceptance.  
Note that the coincidence requirement of two electrons 
\underline{without any restriction on energy}
provides an acceptance approximating that shown in 
Fig.~\ref{fig:4-6-egam-acc-from-origin}.  
This description, as discussed more fully in Sec.~\ref{sec:system_operation}, 
permits luminosity 
to be measured which only indirectly depends on energy calibration.  
At the same time, consistent comparisons of the relevant data distributions 
with predictions validate the calibration. 

The actual acceptance, however, varies slightly from the curve shown 
in Fig.~\ref{fig:4-6-egam-acc-from-origin} 
because the photon beam's finite extent in $y$ creates 
complications.  Photons displaced from the beam centroid shifts the 
contours in Fig.~\ref{fig:4-5-z-vs-egam-contours} 
by amounts of order $10\%$ for each centimeter 
of $y$-displacement. In addition, though the calorimeters sample the 
full transverse range of the transmitted beam, the correlations between 
vertical beam coordinate, photon energy, and acceptance require that we 
address the beam properties in the acceptance calculations.  Full acceptance 
calculations include the entire spectrometer setup, the measured beam shape, 
and the complete shower development in the calorimeters as described 
in Sec.~\ref{sec:system_operation}.  

\subsection{Small Corrections and Uncertainties}
The magnet current was chosen so that, during normal operation, 
the data cover the flat part of the bremsstrahlung energy spectrum 
in Fig.~\ref{fig:4-1-betaheitler-new}.  This ensures that the sensitivities 
to energy scales and alignments are minimized.  
Certain additional, small corrections were included, as described here.

The measured rate requires correction for backgrounds from electron 
beam interactions with the residual gas in the beam pipe. 
This is determined by utilizing the measured coincidence rates 
corresponding to pilot bunches with electrons unpaired with proton bunches. 
For typical running conditions this effect on the luminosity measurement 
is $\sim 1\%$, and the uncertainty is much less.  
Proton beam related backgrounds 
are negligible. Correction for multiple photons observed in a single 
bunch crossing is also negligible. The overall precision of the 
luminosity measurement is discussed more 
fully in Sec.~\ref{sec:system_operation}

\section{System Operation}
\label{sec:system_operation}

Luminosity measurements provided by the spectrometer system utilize spectra 
of the bremsstrahlung photon position and energy from the calorimeters.  
$In$ $situ$ procedures allowed for continuous validation of 
the precisions for 
these quantities.  Specifying accurately (a) the axes of the beam coordinate 
system relative to the calorimeter modules; and (b) the relative and absolute 
energy measurements by the calorimeters were necessary in circumstances 
in which these changed over time.  These issues are discussed in turn.

\subsection{Position Measurements}
\label{sec:pos_measurement}
The photon position is derived directly from the measurements of the 
individual electrons as specified in Eqns.~\ref{eqno6} and \ref{eqno7}.  
Measurements in the two transverse directions have somewhat different 
precisions. 

\begin{figure}[ht]
\begin{center}
\includegraphics[width=8.0cm]{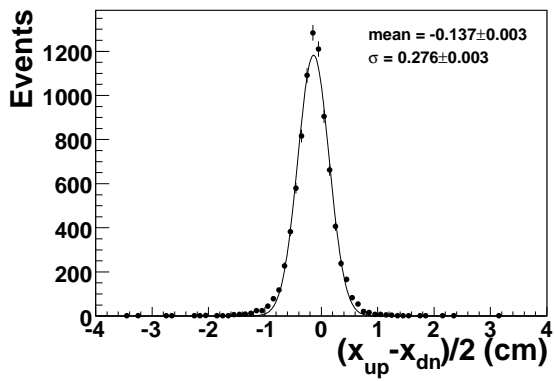}
\caption{Half the difference between $x$-coordinates measured in up and down 
calorimeters. This is a direct measure of the resolutions on $x_{\gamma}$.}
\label{fig:5-1-xres}
\end{center}
\end{figure}

Because electrons are deflected by the magnetic field primarily in the 
vertical plane, the horizontal positions in the upper and lower calorimeters 
should be nearly equal.  The average (Eqn.~\ref{eqno6}) is used to estimate 
the value of $x_{\gamma}$.  Fig.~\ref{fig:5-1-xres} 
shows the complementary difference
between up and down measurements $[(x_{up}-x_{dn})/2 ]$.  The expected mean 
of this quantity should be nearly zero and the standard deviation should be 
similar to the error on $x_{\gamma}$.  The observed RMS of about 3 mm is 
close to the anticipated value, and is dominated by the scattering of the 
electrons in the exit window.  The small offset of the mean is attributed 
to uncorrected vertical components of the magnetic field and a small horizontal
misalignment of the calorimeter modules.  

\begin{figure}[ht]
\begin{center}
\includegraphics[width=8.0cm]{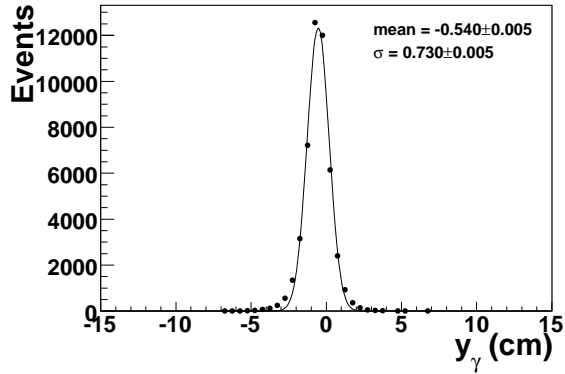}
\caption{The spectrum of reconstructed $y_{\gamma}$ positions with the slit 
collimator inserted into the beamline. Note that the mean of the spectrum 
is consistent with the offset of the slit collimator centroid.}
\label{fig:5-2-yres}
\end{center}
\end{figure}

Figure~\ref{fig:5-2-yres}, the spectrum of 
reconstructed $y_{\gamma}$ coordinates 
obtained with the 2 mm high slit collimator inserted, provides a rough 
estimate of the RMS for measurements of $y_{\gamma}$.  Note that this 
measurement is compromised by the correlations with the energy measurement 
reflected in Eqn.~\ref{eqno7}, but also by the finite width of the slit, 
and by scattering from the slit edges.  Hence, the RMS of 7 mm should be 
considered an upper limit on the actual precision on $y_{\gamma}$.  
The edge scattering 
also is known to produce the tails in the distribution of 
Fig.~\ref{fig:5-2-yres}.

\begin{figure}[ht]
\begin{center}
\includegraphics[width=8.0cm]{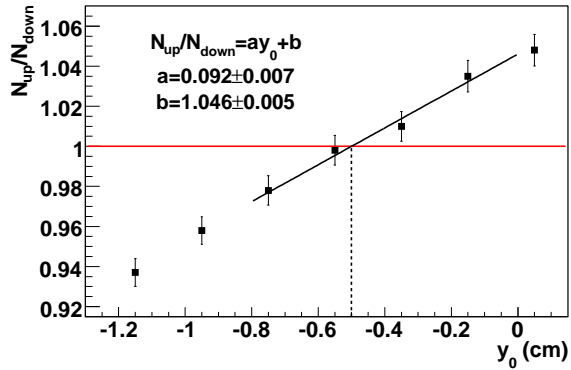}
\caption{Ratio of single electrons observed in the up calorimeter ($N_{up}$) 
to that in the down calorimeter ($N_{dn}$) as a function of the assumed offset 
of the collimator center ($y_{0}$).}
\label{fig:5-3-nup-ndn}
\end{center}
\end{figure}

Because the calibration of the calorimeter energy scale depends on vertical 
alignment, as described in Sec.~\ref{sec:luminosity_calculation}, 
a technique for validating the vertical collimator axis origin was important.  
Accomplishing this utilized the fact that the rates in upper and lower 
calorimeter modules must be equal at equal vertical distances from the 
beam centroid.  Figure~\ref{fig:5-3-nup-ndn}, the ratio of electron rate in 
up and down detectors while the slit collimator was inserted versus $y_{0}$, 
the offset of the collimator centroid, illustrates the determination 
of the correct offset.  From the figure, we see that the offset of the 
collimator midpoint was located at $y_{0}=-0.50$~cm. This value is consistent 
within error with the value taken from the optical survey, which is 
$y_{0}=-0.56$~cm.

\begin{figure}[ht]
\begin{center}
\includegraphics[width=8.0cm]{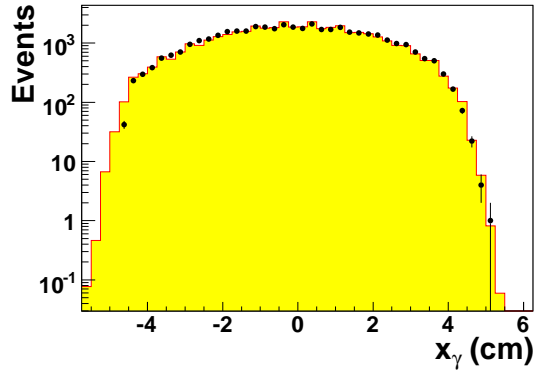}
\caption{The spectrum of reconstructed $x_{\gamma}$ positions for accepted 
photons shown on a logarithmic scale. The Monte Carlo predictions are shown 
as the histogram. Note the high statistics of the data, which were collected 
in one 16 second period.}
\label{fig:5-4-xlog}
\end{center}
\end{figure}

\begin{figure}[ht]
\begin{center}
\includegraphics[width=8.0cm]{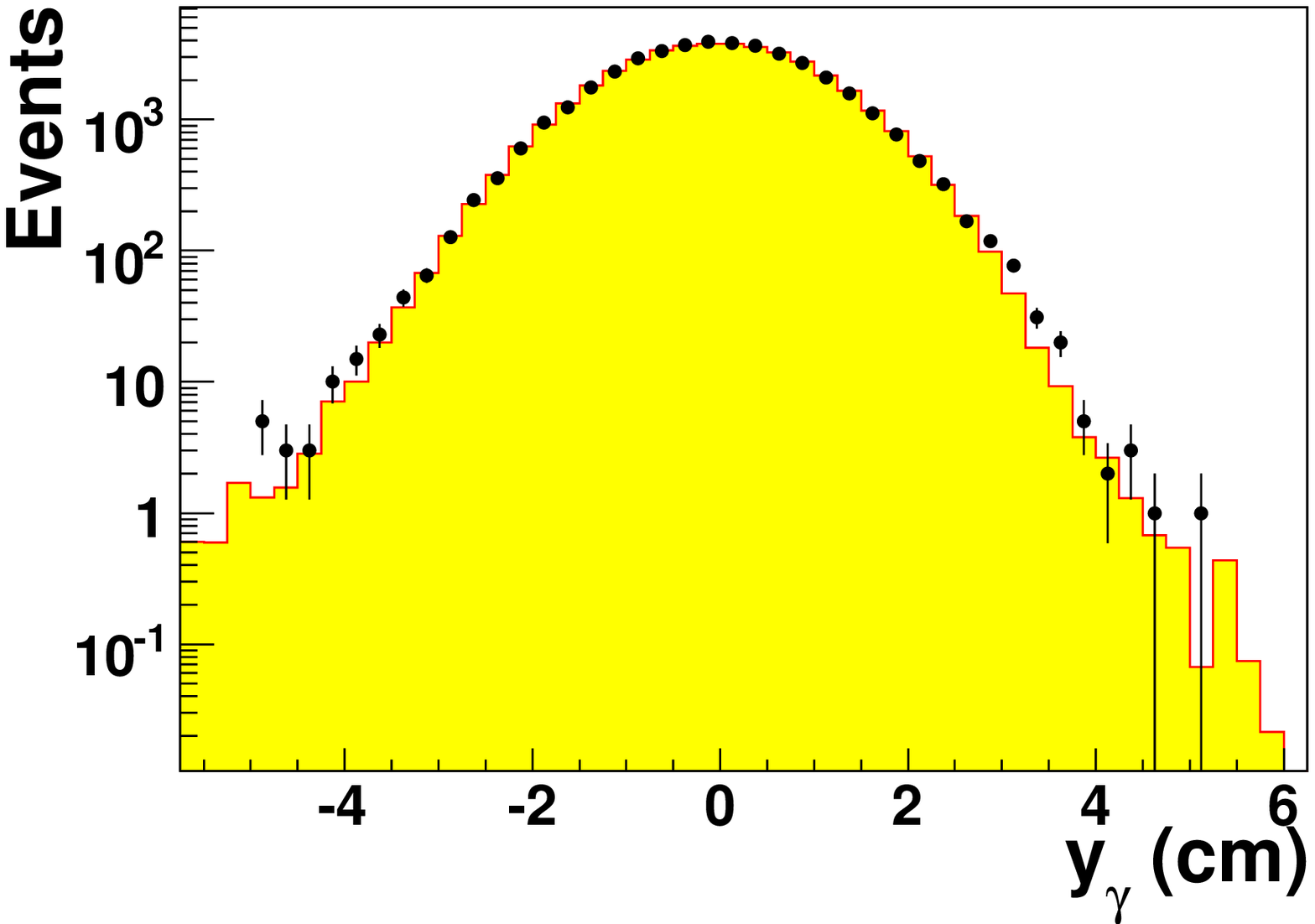}
\caption{The spectrum of reconstructed $y_{\gamma}$ positions for accepted 
photons shown on a logarithmic scale. The Monte Carlo predictions are shown 
as the histogram. Note the high statistics of the data, which were collected 
in one 16 second period.}
\label{fig:5-5-ylog}
\end{center}
\end{figure}

Examples of projected beam profiles for $x_{\gamma}$ and $y_{\gamma}$ are 
shown in Fig.~\ref{fig:5-4-xlog} and Fig.~\ref{fig:5-5-ylog}, respectively, 
on logarithmic scales. The Monte Carlo simulation (histogram) reproduces 
the measured data (points) well, even on the tails where the rate has fallen 
significantly.  The plots justify the use of single Gaussians fits to 
the beam profiles in the Monte Carlo calculation providing the acceptance. 

It is possible for the beam to continually change mean positions and widths 
by small amounts in either dimension.  Hence measurements of beam profiles
for $x_{\gamma}$ and $y_{\gamma}$ are recorded every 16 seconds 
\footnote{This sampling time is more than one order of magnitude smaller 
than the typical beam polarization time, and substantially smaller than 
typical changes in the beam profile.},
so that corrections for the upstream aperture 
restrictions and beam parameters were made continuously to the 
luminosity calculation.

\subsection{Energy Measurements}
The individual electron energies, from a bremsstrahlung photon conversion, 
are related to their vertical displacements at the calorimeter relative to 
the photon position (Eqn.~\ref{eqno12}).  
When the slit collimator is inserted, the distance of the
reconstructed electron from the collimator opening (as determined
in Sec.~\ref{sec:pos_measurement} and Fig.~\ref{fig:5-3-nup-ndn}), 
along with the $\int B dz$ from the
field map, provide an accurate measure of the electron energy.
Data taken in this configuration were used to determine the
individual channel gains $S_i$ in Eqn.~\ref{eqno1}.
Because the position reconstruction depends weakly on the values
of the $S_i$ as seen in Eqn.~\ref{eqno3}, the calibration is repeated
using the new gains; the procedure converges after a small number
of iterations.

\begin{figure}[ht]
\begin{center}
\includegraphics[width=8.0cm]{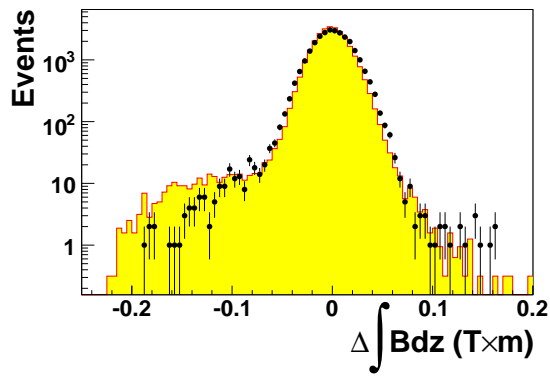}
\caption{The spectrum obtained from observed $e^{+}e^{-}$ conversions for the 
integrated field calculated from the events' measured energies and positions 
minus the 
nominal integrated field. The nominal value is $\int B dz \sim 0.32$ Tm.}
\label{fig:5-6-dbdl}
\end{center}
\end{figure}

A consistency check between the calibration and magnetic field
map was also performed. If the calorimeter 
calibrations are well understood, each event permits an evaluation of both 
the event $y_{\gamma}$ and the value of the integrated magnetic field 
traversed by the pair.  We define $\Delta \int B dz$ as the integrated field 
measured from the electron energies and positions minus the value known from 
the measured magnetic field map.  The distribution in $\Delta \int B dz$ 
for typical operating conditions is shown (data points) in 
Fig~\ref{fig:5-6-dbdl}, 
and the histogram is predicted by the Monte Carlo simulation.  The RMS width 
is about $10\%$ of the typical $\int B dz \sim 0.32$ Tm, which is 
consistent with the individual electron energy resolution described in 
Sec.~\ref{sec:spectrometer_design_and_components}.  The asymmetric tail to
negative $\Delta \int B dz$ is attributed to the small number ($\sim 1\%$) 
of photons converting inside the magnet aperture in comparison to 
those converting in the exit window.  

The overconstraints in the reconstruction were very important and used 
frequently, particularly during early operations when severe background 
conditions caused some deterioration of calorimeter response because 
of radiation damage to the calorimeter WLSs.  
This issue will be discussed later in this section.

\begin{figure}[ht]
\begin{center}
\includegraphics[width=8.0cm]{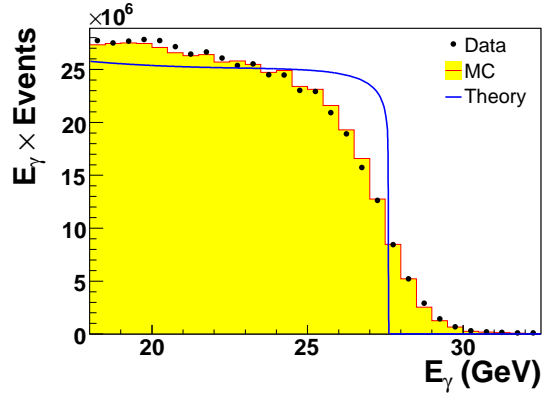}
\caption{The observed spectrum corrected for acceptance 
(multiplied by $E_{\gamma}$) as a function of $E_{\gamma}$ for the magnet 
current raised so as to observe the spectrum end point. The continuous curve 
specifies the spectrum shape predicted for a perfect detector. The histogram 
shows the Monte Carlo prediction for this detector.}
\label{fig:5-7-endpoint}
\end{center}
\end{figure}

One additional independent check of the energy scale was performed --- a test 
against the known circulating electron beam energy.  By increasing the dipole 
magnetic field and acquiring data, a value was empirically determined 
for the end point of the bremsstrahlung spectrum with our 
calibrations of calorimeter and magnet.  The data for one such run, 
after correction for acceptance, are shown in Fig.~\ref{fig:5-7-endpoint}.  
The curve shows the Bethe-Heitler spectrum, multiplied by the photon energy,
in the vicinity of the end point. The points show the data, corrected 
for acceptance. The shaded histogram shows the predicted spectrum for this
quantity from the Monte Carlo, with energy adjusted to best fit the observed 
data. The
adjusted energy is within $1\%$ of the expected value, 27.6~GeV, which is
the known energy of the electron beam.

\subsection{Radiation Complications}
\label{sec:radiation}
Under operating conditions, the most severe unanticipated complications 
arose due to severe radiation backgrounds in the region of the calorimeters.
During beam tuning, and even during stable operations, the synchrotron 
radiation in the region of the calorimeters was high, particularly in the 
scintillator strips near the beamline.  The wavelength-shifting light guides 
in the regions of the PMTs were particularly susceptible to radiation damage.  
Under severe conditions, the calibrations of some channels would change, 
and then would later recover during periods when the beam was absent.  
These effects were discovered because of the available overconstraints 
that permitted $in$ $situ$ calibrations.  Corrections for the radiation damage 
maintained the scale within $5\%$ of the nominal, and was understood 
well enough to correct for small triggering threshold effects.  
Hence, the issue was managed by performing periodic calibrations and 
utilizing these to make appropriate corrections.

\begin{figure}[ht]
\begin{center}
\includegraphics[width=12.0cm]{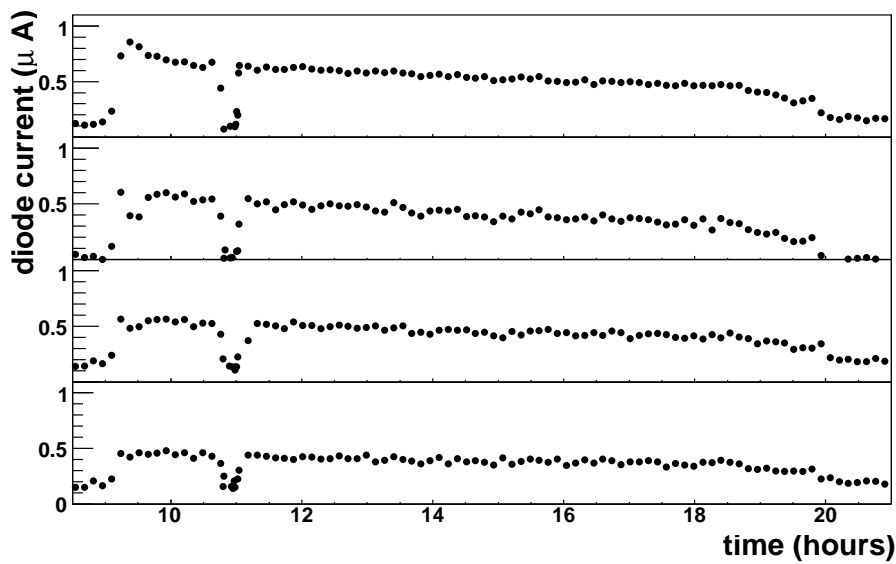}
\caption{Typical data from four radiation monitors versus time. Beam 
acceleration and collimator insertion are clearly visible (see text).}
\label{fig:5-8-diodes}
\end{center}
\end{figure}

After diagnosis, the operational effects of the radiation field were reduced 
by making 
physical modifications during shutdowns.  Additional, well-designed shielding 
was installed in the 
affected regions.  In the process, radiation monitors 
were also installed in order to monitor the severity of the radiation 
backgrounds and to identify the regions of damage.  Such monitors proved to be 
important because seemingly small changes in beam configurations could result 
in substantially different radiation fields at the detector.  
Large changes in beam conditions, like switching from circulating positron 
beam to electron beam, produced even larger effects.  
Figure~\ref{fig:5-8-diodes} 
provides a typical example of data taken with four different radiation 
monitors versus time.  In these figures, the electron beam reaches full 
energy at 9.0 hours, and the collimator blocking 
the bremsstrahlung (and synchrotron) beam was inserted at 10.7 hours, 
then removed at 11.2 hours.  The presence of the large radiation 
field is clearly seen to originate from the beamline.

During early running, when the radiation problem was first discovered,
and with the hardware trigger level set to a relatively large value,
the channel gain drifts because of radiation damage 
contributed a $2-3\%$ systematic error on the acceptance calculation because
of good events not firing the hardware trigger, as estimated by Monte
Carlo simulation. Another problem associated with channel gain drifts is the
$y$ position measurement of the photon, which is related to the 
energy measurement of each calorimeter as shown in Eqn.~\ref{eqno7}.
Subsequently, the error in the acceptance calculation was reduced
to a negligible level by employing frequent software calibrations.

\subsection{Bremsstrahlung and Pair Production Measurements}
\label{sec:brem_select}
Good bremsstrahlung photons are selected from electron-positron coincidences 
at the two calorimeters in a two-step process. The first stage,
employed by the hardware trigger, requires both calorimeters to have energy
deposits exceeding a certain threshold, as mentioned in Sec.~\ref{sec:daq}.
The small threshold value is chosen to distinguish real electrons 
from electronic noise. According to the
geometry setup of the system, the energy spectrum for electrons 
from pair production of the bremsstrahlung photon begins to rise from zero 
around 6~GeV, while the total noise level
for the calorimeters is lower than 1~GeV under normal running 
circumstances.
However, since the calibration constants are only applied later in software,
the rising edge of a good electron spectrum can have tails, and depends on
the hit position because some channels are more 
radiation exposed than others.
Cutting out the tail with the hardware trigger would add error to the 
luminosity calculation.
The tails were minimized by periodic high-voltage
trimming of the detectors, and by choosing a small hardware threshold which is 
far away from the noise level.

For events with signals from both detectors accepted by the hardware trigger, 
a software
selection including the following requirements is applied to both detectors:
\begin{itemize}
\item{the reconstructed energy for each detector must be larger than 3.5~GeV, 
which guarantees the coincidence requirement;}
\item{the standard deviation of shower strip positions (obtained using an 
algorithm with linear energy weighting) 
must be less than 1.0~cm, distinguishing electromagnetic showers from hadronic 
showers;}
\item{the largest energy depositions in both $x$ and $y$ directions are not at 
the edge strips, guaranteeing good energy and position reconstruction.}
\end{itemize}
Also, a requirement on the reconstructed $y$ position in the down detector 
electron is imposed to make sure the fiducial regions for up and down 
detectors are
symmetrical with respect to the $y$ position defined by the moving collimator.

The coincidence selection requirements lower the 
background to a negligible level. With the good bremsstrahlung photons 
selected,
the photon physics quantities (energy, $x$ and $y$ positions)
can be reconstructed as described in
Sec.~\ref{sec:recon_photon}. Below we show some of the important physics 
quantities compared with Monte Carlo predictions. 
 
\begin{figure}[ht]
\begin{center}
\includegraphics[width=8.0cm]{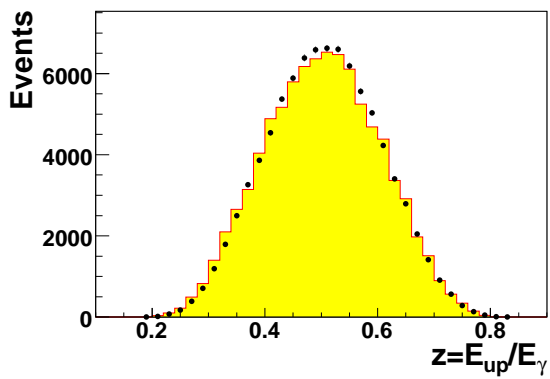}
\caption{An observed $z=E_{up}/E_{\gamma}$ spectrum with the Monte Carlo 
prediction (histogram).}
\label{fig:5-9-z}
\end{center}
\end{figure}

As described in Sec.~\ref{sec:luminosity_calculation}, one critical 
underlying physics process for the luminosity measurement involves 
the $z$-distribution reflecting the energy sharing in the pair production 
process, shown in Fig.~\ref{fig:4-2-raw-z-dist}. We show in 
Fig.~\ref{fig:5-9-z} the reconstructed spectrum under typical running 
conditions compared with the Monte Carlo simulation (histogram) 
incorporating the acceptance, resolutions, and calibrations.  
The good agreement indicates that measurements are well understood and 
simulated properly.

\begin{figure}[ht]
\begin{center}
\includegraphics[width=8.0cm]{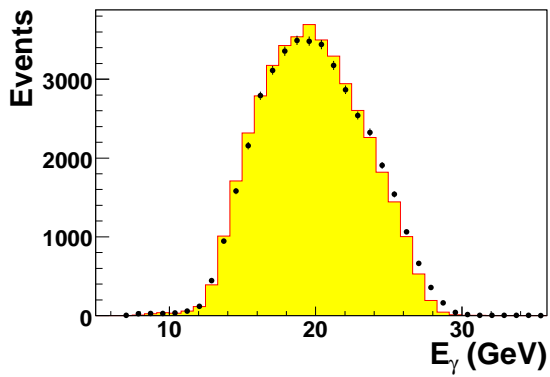}
\caption{An observed $E_{\gamma}$ spectrum with the Monte Carlo prediction 
(histogram). Note the high statistics of data, which were collected in one 
16 second period.}
\label{fig:5-10-egam}
\end{center}
\end{figure}

A second critical process is the bremsstrahlung creation, whose cross 
section is shown in Fig.~\ref{fig:4-1-betaheitler-new}.  
The distribution of observed photon energies, shown in 
Fig.~\ref{fig:5-10-egam}, 
is compared with the Monte Carlo simulation.  Again, the 
agreement is 
good and illustrates that the 
instrument is well understood. Note that the 
acceptance is zero at the low values 
of $E_{\gamma}$ and at both low and high values of $z$, 
so that electrons of low energy
$E_{e} \sim 3.5$~GeV (corresponding to, for example, 
$z=0.3$ and $E_{\gamma}=12$~GeV) have very low acceptance by virtue of 
geometry and the magnetic field. 
In other words, for both $E_{\gamma}$ and $z$ to be observed with small 
values, the beam $y$ 
position must be off center by a large amount,
and this is carefully avoided during normal HERA operation.

\subsection{Measurements of Luminosity}
\label{sec:lumi_calculation}
The integrated luminosity is calculated based on the 
number of good bremsstrahlung 
photons selected as
described in Sec.~\ref{sec:brem_select}. Two important issues in this 
calculation need
to be addressed: deadtime and acceptance.

For the trigger setting of $\sim 2$~GeV at both detectors 
and the prescale set to 1, 
the DAQ system deadtime ranged from about $80\%$ at the start of a HERA fill 
to $20\%$ at the end. With the typical operating prescale at 3, 
the deadtime was reduced by more than a factor of two at the start of fill and
usually dropped to essentially zero later in the fill.  
Differences between results from the 
two methods of deadtime estimation mentioned in Sec.~\ref{sec:daq} 
were smaller than one percent 
in most cases, and consistent with statistical fluctuations.

The typical data rate under the current trigger setting and prescale factor 3 
is around 3~kHz. After software selections the data rate drops by about 
a factor of 2.
Still, it is neither feasible nor necessary to retain an individual data 
record for 
every good bremsstrahlung photon.
The following data are saved every 16 seconds for both the offline 
luminosity calculation and beam monitoring purposes:
\begin{itemize}
\item{histograms for $x$, $y$ positions and energy of the bremsstrahlung 
photons;}
\item{counters for coincidences in each HERA bunch;}
\item{counters for total HERA bunches during the period;}
\item{counters for HERA bunches when the DAQ system buffer is not full 
(actively receiving data);}
\item{correlator between $x$, $y$ position of bremsstrahlung photons, 
$<x_{\gamma}y_{\gamma}>$.}
\end{itemize}

A Geant~3.21 simulation of the entire spectrometer system is used in the
offline acceptance calculation. This simulation incorporates the measured 
aperture,
exit window, dipole and detectors. Additional conversions in air are simulated,
as is scattering of the $e^{\pm}$ in the window and air.

First, the beam ellipse tilt is calculated using the moments of the 
$x_{\gamma}$ and $y_{\gamma}$ histograms and the correlator 
$<x_{\gamma}y_{\gamma}>$. Using this tilt, a sample of simulated events is 
reweighted and fit to the $x_{\gamma}$ and $y_{\gamma}$ histograms. The
result of this fit are the photon beam horizontal and vertical mean positions 
and 
RMS spreads, along with the acceptance corrected number of bremsstrahlung 
events. 
This acceptance corrected number of events is then used to calculate the 
luminosity using the known bremsstrahlung cross section. 
An example of
the output of the fit was shown in Figs.~\ref{fig:5-4-xlog} 
and \ref{fig:5-5-ylog}. 
Note the high statistics accumulated in the 16 second integration period.
The $E_{\gamma}$ 
histogram, in Fig.~\ref{fig:5-10-egam}, agrees well with the
Monte Carlo prediction without any additional parameters. 

Some small additional corrections must be applied to obtain the final 
luminosity, such as
good coincidences not firing the hardware trigger before the trigger 
threshold
was lowered. The trigger is simulated
by shifting the gain of each channel in Monte Carlo by the amount observed
in the most recent calibration. This is typically a $2-3\%$ correction. 
Another concern caused by the gain drifts from
radiation are systematic shifts in the 
$y$ position of the photon.
This was tested by Monte Carlo with the maximum observed gain drifts.
The effect is found to be at most a few per cent and completely negligible
when frequent calibrations are being carried out. 
During normal running conditions,
a calibration is done at the end of each fill of HERA.

\begin{figure}[ht]
\begin{center}
\includegraphics[width=12.0cm]{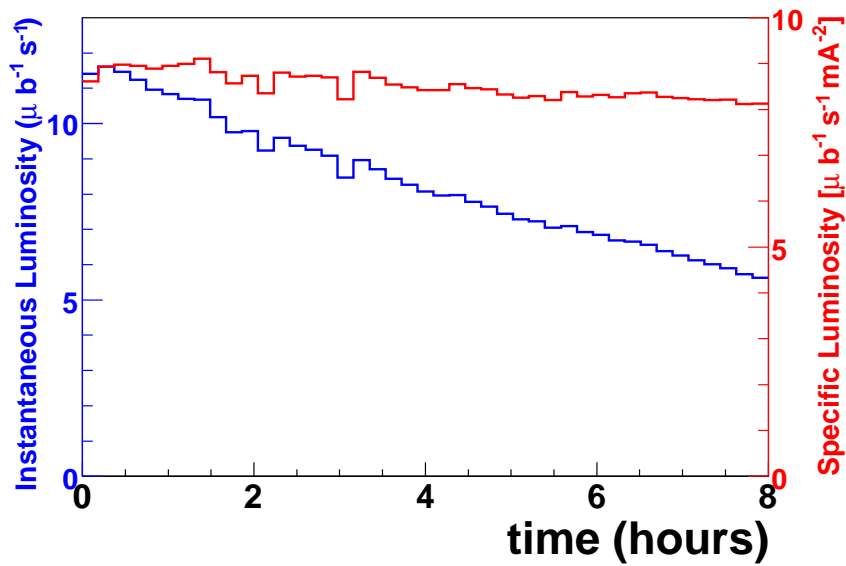}
\caption{Lower histogram shows the instantaneous luminosity (left scale); 
the upper histogram shows the specific luminosity (right scale), during a 
typical beam fill. Note that the luminosity decrease during the fill 
primarily arises from loss of electron current.}
\label{fig:5-11-luminosity}
\end{center}
\end{figure}

The calculated luminosities for a typical 8-hour run are shown 
in Fig.~\ref{fig:5-11-luminosity}.  
The lower curve, with values shown on the left hand axis, is the instantaneous 
luminosity reported by the monitor.  The decrease in luminosity during the 
period is primarily attributable to the decrease in electron beam current 
over the period.  The upper curve in the figure, with values shown on the 
axis to the right, is the instantaneous specific luminosity, obtained by 
dividing by the sum of the product of proton and electron bunch currents as in
Eqn.~\ref{eqno9}. Note that the
instantaneous specific luminosity decreases much more slowly, which is due to
the increase of the proton beam emittance.

\subsection{Systematic uncertainties}
The systematic uncertainties from various sources 
are given in Tab.~\ref{tab:uncertainties}.
\begin{table}[ht]
\begin{center}
\begin{tabular}{|c|c|}
\hline
\textbf{Cause} & \textbf{Uncertainty in Luminosity} \\
\hline
Vertical alignment and $y_{\gamma}$ measurement & 2.5\% \\
\hline
Photon conversion rate & 2\% \\
\hline
Pile-up &  0.5\% \\
\hline
Deadtime measurement & 0.5\% \\
\hline
Theoretical Bethe-Heitler cross section & 0.5\% \\
\hline
Dipole magnetic field & small \\
\hline
Trigger threshold correction & small \\
\hline
\hline
Total & 3.5\% \\
\hline
\end{tabular}
\label{tab:uncertainties}
\caption{Estimated systematic errors associated with the issues creating 
the largest effects.}
\end{center}
\end{table}

The shape of the aperture and its alignment relative to the detectors
were each measured to an accuracy of 1 mm. The effects of gain drifts could
cause a shift in $y_{\gamma}$ measurement of 2 mm. To check the effect on the
acceptance, a Monte Carlo simulation of the photon beam was moved by these
amounts. The overall uncertainty on the acceptance from these effects is
$2.5 \%$, as given in the first row of the table.

The acceptance is directly proportional to the fraction of photons converted
in the exit window (a very small faction from the air also). 
Uncertainties on the cross section for conversion, 
material in the window, and the window thickness lead to an uncertainty on the
acceptance of $2\%$, shown in the next row.

Several other effects were considered and estimated to be $<0.5\%$. These
include pile-up of bremsstrahlung photons, deadtime measurement, knowledge
of the dipole field, trigger threshold correction, and the theoretical
Bethe-Heitler cross section.

For the early operations, uncertainties in the acceptance, window
conversions, and other issues resulted in an overall $3.5\%$ estimated error in
the luminosity determination. With time and experience, this error will
decrease. The most important step 
in achieving this will be the direct and empirical
determination of the acceptance and conversion fraction, using the 6 meter
tagger located near the IP.  This device collects an unbiased sample of
recoil bremsstrahlung electrons, with companion photon energies within the
acceptance of the luminosity spectrometer.  For such tagger events, the 
fraction
with detected photons constitutes a direct measure of the product $fA$ in
Eqn.~\ref{eqno10}.  After appropriate experience and experimentation, we
anticipate the ultimate uncertainty in luminosity measurement to be about
$2\%$.

\section{Summary}
In summary, the luminosity spectrometer has been successfully brought into
operation.  The device has been used to measure luminosity for the 2004 data
collection by ZEUS with an estimated error of $3.5\%$.  Operations were largely
as anticipated.  The only major obstacle was the very intense synchrotron
radiation field in the region of the calorimeters.  This radiation
compromised somewhat the operations of the wavelength-shifting readout of
the calorimeter scintillators.  With recent and ongoing shielding
improvements, along with direct measurements of the product of the photon
conversion fraction and acceptance, we anticipate achieving luminosity
measurements with about $2\%$ uncertainty 
with the luminosity spectrometer system.

\section*{Acknowledgments}
We thank I.~Abt for her help on the construction of the radiation monitor 
and her very helpful discussions and comments on the paper, 
M.~Gil and W.~Ruchlewicz for their support on the DAQ system, 
C.~Muhl for his help on the detector shielding construction,
and U.~Stoesslein for her help on ZEUS offline 
software organization.

\end{document}